\documentclass[aps,showpacs]{revtex4}
\usepackage{graphicx,epsf}%

\makeatletter
\def\btt#1{\texttt{\@backslashchar#1}}%
\DeclareRobustCommand\bblash{\btt{\@backslashchar}}%
\makeatother
\def \slash#1{\not\! #1}

\begin{document}
\large
\title[]{About one method of analytical calculations of reaction amplitudes with fermions}
\author{Victor Andreev}
\email{andreev@gsu.unibel.by}
\affiliation{Gomel State University, Physics Department\\
246699 Gomel, Belarus\\
}

\date{\today}

\begin{abstract}
The paper presents the new method of calculations of reaction
amplitudes of processes with spin $1/2$ fermions. The method is
based on the application of isotropic tetrad in Minkowski space
and basis spinors connected with it. We obtained as test and
illustration of the method the amplitudes of interaction
processes $e^-e^+\to f \bar{f} $, $e^- e^+\to W^-W^+ $ and one of
the possible diagrams of the process $e^+e^-\to
e^+e^-e^+e^-e^+e^-$.
\end{abstract}


\maketitle

\section{Introduction}
\label{sec:level1}

Now a comparison the consequences of quantum field theories with
the results of experiments in a high-energy physics requires
calculation of observed quantities with a high scale of
precision. The standard way to obtain a cross section with the
fermions of a spin $1/2$ in perturbative quantum field theories
is to reduce the squared amplitude to a trace from products of
$\gamma$-matrices. The study of multi-particle  final states in
reactions and the investigation of polarization effects have
required a new effective method of calculations, because the
standard method involves the calculation of a great number of
traces with a lot of Dirac $\gamma$-matrices.

The method of direct calculation of matrix elements became the
alternative to the standard method. The idea of calculating
amplitudes has a long enough history. In 1949  it was suggested
in \cite{Powell} to calculate a matrix element by means of
explicit form of $\gamma$-matrices and Dirac bispinors (more
detailed bibliography on the problem can be found in
\cite{Galynski,Bondarev}).

Now there have been worked out a lot of methods of calculating the
reaction amplitudes with fermions of the spin $1/2$. That's why
one can say that the method is becoming a standard method in
high-energy physics for obtaining cross sections and decay rates.

The methods of matrix elements calculations can be divided into
two major classes. The first class includes the methods of direct
numerical calculation of the Feynman diagrams (see, for example,
\cite{Farrar}). The second class includes the methods of
analytical calculations of amplitudes with the subsequent
numerical calculation of cross sections. It should be noted that
there are methods of calculating cross sections without the
Feynman diagrams \cite{Giele}-\cite{Helac}.

The analytical methods of Feynman amplitudes calculation can be
divided into two groups. The first group involves the analytical
methods that reduce the calculation of  $S$-matrix element to
trace calculation. The procedure of trace calculation underlies a
lot of methods (see, \cite{Galynski,Bondarev},
\cite{Bellomo}-\cite{Vega} etc.). In the methods matrix element is
reduced to the combination of scalar products of four-vectors and
their contraction with the Levi-Civita tensor.

The second group involves the analytical methods that practically
do not use the operations with traces from products of
$\gamma$-matrices. The method of the CALCUL group which was used
for the calculations of the reactions with massless fermions is
the most famous among \cite{Berendz}-\cite{Zhan}. In this method
the matrix element is reduced to spinor products of bispinors i.e.
$\overline{U}_{\lambda}\left(p\right)U_{-\lambda}\left(k\right)$.
The spinor products are calculated through momentum components by
means of traces. However, the operation of matrix element
reduction is not so simple as the calculation of traces. It
requires the use of Chisholm spinor identities (see
\cite{Kleiss}). Also it takes the representation of contraction
$\slash{p}=p^\mu \gamma_\mu$ with four-momenta $p^\mu$ and
polarization vectors of external photons through bispinors. For
gauge massive bosons the additional mathematical constructions
are needed \cite{Kleiss}.

There are generalizations of the method for massive Dirac
particles both for special choices of the fermion polarization
(\cite{Kleiss,Kleiss1,Hagiwara}) and for an arbitrary fermion
polarization \cite{Gongora,Andreev}. We call the polarization
states of fermions in Ref.\cite{Kleiss} as Kleiss-Stirling or
$KS$-states.

In Ref.~\cite{Hagiwara} the original algorithm of reduction to
spinor products for helicity massive fermions with the help of the
Weyl representation for $\gamma$-matrices  was offered. It should
be noted that for massless fermions we can obtain amplitude in
terms of the scalar products of four-momentum vectors and
current-like constructions of the type
$J^{\mu}\sim\overline{U}_{\lambda}\left(p\right)\gamma^{\mu} U _
{\lambda} \left (k\right)$. The components of $J^{\mu}$ are
calculated through momentum components $p,k$ (so called $E$-vector
formalism, see \cite{Papadopoulos}).

For $KS$-states Ref.\cite{Ballestrero} presents the iterative
scheme of calculation that reduces expression for the fermion line
$\overline{U}_{\lambda} \left(p\right)Q U _{\lambda}
\left(k\right) $ to the combination of spinor products
$\overline{U}_{\lambda}\left(p\right)U_{\lambda}\left(k\right)$
and (or)
$\overline{U}_{\lambda}\left(p\right)\gamma^{\mu}\left(g_V+g_A
\gamma_5 \right) U_{\lambda} \left(k\right)$ by means of inserting
the complete set of  non-physical states of bispinors (with
$p^2<0$) into the fermion line.

It goes without saying that  both the spinor products and current
constructions in all the methods were calculated by means of
traces and then used as a universal function similar to the
expression of scalar product of four-vectors through their
components.

It is very difficult to estimate the efficiency of different
methods of matrix element calculation, as their application often
comes from the physical problem that must be solved. So, for
example, the methods of calculation by means of traces are
universal and they don't require any additional constructions of
polarization vectors of particles as in spinor techniques.
However, in the trace method the number of terms to be evaluated
increases even quadratically with the number of diagrams.

The efficiency of spinor techniques is that the evaluation of
matrix element is expressed through the spinor products that have
been calculated before. It decreases the number of terms and it is
convenient when we pass over to numerical calculations
immediately. However, simple process $e^+e^-\to W^+W^-$ with
polarized $W$-bosons for spinor technique is more difficult to
calculate than it is in trace method.

The disadvantages of other methods are that they use a specific
choice of polarization vectors and accordingly require additional
calculations if it is necessary to calculate other polarization
configurations of fermions.

Many analytical methods underlie both  universal programs of
matrix element calculations and cross sections $CompHEP$
\cite{Comphep}, $GRACE$ \cite{Grace}, $FeynCalc$ \cite{FeynCalc},
(see also \cite{Stelzer}) and special-purpose programs. The
detailed list of such programs can be found in
Ref.\cite{Harlander}.

The aim of this paper is to present a new method for calculating
the amplitudes of processes involving both massive fermions of an
arbitrary polarization and massless fermions. This method is based
on the use of an isotropic tetrad in Minkowski space and basis
spinors connected with it. Here we don't use  an explicit form of
bispinors and $\gamma$-matrices or operation of trace
calculations.

In this method as well as in the trace methods the matrix element
of Feynman amplitudes is reduced to the combination of scalar
products of momenta and polarization vectors. Unlike spinor
technique in different variants \cite{Berendz}-\cite{Kleiss1},
this method doesn't use either Chisholm identities, or the
presentation of the contraction $\slash{p}$ with four vector $p$
and of the polarization vector of bosons through the bispinors.
The minimum number of operations and the simple algorithm of the
method make it very efficient and not complicated in calculations.

As the suggested method is based on the active use of basis
spinors connected with isotropic tetrad vectors we shall call it
the method of basis spinors (MBS).

The structure of the paper is as follows: in Sec.~\ref{sec:level2}
we define an isotropic tetrad and complete set of spinors
connected with it. We give the main formulas that underlie the
suggested method. In Sec.~\ref{sec:level3}  we consider the
decomposition coefficients of bispinors with an arbitrary
polarization vector on basis spinors. We give decomposition
coefficients for fermion with the most frequently used
polarization states. In Sec.~\ref{sec:level4} we  render the
method of basis spinors and give a brief comparison of the
suggested scheme of calculations with methods of spinor
techniques and trace method. Sec.~\ref{sec:level5} contains the
calculation of the amplitudes of the reaction $e^-e^+\to f \bar
f$ with massive fermions $f$, the process $e^-e^+\to W^-W^+$ with
polarized $W$-bosons and one of possible Feynman diagrams of
reaction $e^+e^-\to e^+e^-e^+e^-e^+e^-$ (for massless fermions)
as the illustration of the MBS. The last section traditionally
contains the concluding remarks. Appendix \ref{sec:level7}
includes some relations for matrix elements with massless
fermions.

\section{ Isotropic tetrad and basis spinors. Some relations}
\label{sec:level2}

We use the metric and matrix convention as in the book by Bjorken
and Drell \cite{Bjorken1} i.e. the Levi-Civita tensor is
determined as $\epsilon_{0 1 2 3}=1$ and the matrix $ \gamma_5=i
\gamma ^{0} \gamma ^{1} \gamma ^{2} \gamma ^{3}$. Let us
introduce the orthonormal four-vector basis in Minkowski space
which satisfies the relations:
\begin{equation}\label{stp1}
l_0^{\mu} \bullet l_0^{\nu} -l_1^{\mu} \bullet l_1^{\nu}
-l_2^{\mu} \bullet l_2^{\nu}-l_3^{\mu} \bullet l_3^{\nu} = g^{\mu
\nu}, ~~l_{0}^2=-l_{1}^2=-l_{2}^2=-l_{3}^2 =1,~~\mu,\nu=0,1,2,3.
\end{equation}
In Eq.(\ref{stp1}) $g$ is the metric tensor. We fix the basis
orientation by the relation
\begin{equation}\label{stp1a}
  \epsilon_{\mu \nu \rho \sigma}l_0^{\mu}l_1^{\nu} l_2^{\rho}
  l_3^{\sigma}=-\epsilon_{0 1 2 3}=-1.
\end{equation}
An arbitrary four-vector $p$ can be written as
\begin{equation}
p = \left( p\cdot l_0 \right) \cdot l_0 -\left(p\cdot l_1 \right)
\cdot l_1 -\left( p\cdot l_2 \right) \cdot l_2-\left( p\cdot l_3
\right) \cdot l_3 ~. \label{stp2}
\end{equation}

With the help of the vectors $l_{A}\left(A=0,1,2,3\right)$ we can
define lightlike vectors, which form the isotropic tetrad in
Minkowski space (about isotropic tetrad, see \cite{Borodulin}):
\begin{equation}
b_\rho =l_0+\rho l_3,n_\lambda =l_1+i\lambda l_2,~~~\rho
,\lambda =\pm 1. \label{stp3}
\end{equation}
From Eqs. (\ref{stp1}), (\ref{stp3}) it follows  that
\begin{equation}
(b_\rho \cdot b_{-\lambda })=2\delta _{\lambda ,\rho
}~,~~(n_\lambda \cdot n_{-\rho })=-2\delta _{\lambda ,\rho
}~,~~\left(b_\rho \cdot n_\lambda \right) =0, \label{stp4}
\end{equation}
\begin{equation}
1/2\sum_{\lambda =-1}^1\left[ b_\lambda ^\mu \bullet
b_{-\lambda}^\nu -n_\lambda ^\mu \bullet n_{-\lambda }^\nu \right]
=g^{\mu \nu}. \label{stp5}
\end{equation}
Therefore, Dirac matrix $\gamma^\mu $ and contracting
$\slash{p}=p_\mu \gamma^\mu$ with four-vector $p_\mu$  can be
written as
\begin{equation}
\gamma ^\mu =1/2\sum_{\lambda =-1}^1\left[\slash{b}_{-\lambda}
b_\lambda ^\mu -\slash{n}_{-\lambda } n_\lambda ^\mu \right],
\label{stp7}
\end{equation}
\begin{equation}
\slash{p} =1/2\sum_{\lambda =-1}^1\left[\left(b_{-\lambda } \cdot
p\right) \slash{b}_\lambda -\left(n_{-\lambda} \cdot p\right)
\slash{n}_\lambda \right] . \label{stp6}
\end{equation}

With the help of isotropic tetrad (\ref{stp3}) we define  {\it
basis spinors} $U_{\lambda}\left(b_{-1}\right)$ and
$U_{\lambda}\left(b_{1}\right)$
\begin{equation}
U_\lambda \left( b_{-1}\right) \overline{U}_\lambda
\left(b_{-1}\right) =\omega _\lambda \slash{b}_{-1}, \label{stp8}
\end{equation}
\begin{equation}
U_\lambda \left(b_{1}\right) \equiv
\frac{\slash{b}_{1}}{2}U_{-\lambda}\left(b_{-1}\right).
\label{stp9}
\end{equation}
\begin{equation}
\omega _\lambda U_\lambda \left(b_{\pm1}\right)= U_\lambda
\left(b_{\pm 1}\right) \label{stp10}
\end{equation}
with $\omega _{\lambda} = 1/2 \hskip 1pt \left( 1+\lambda
\gamma_5\right)$.

If we introduce raising and lowering spin operators
\begin{equation}
\frac \lambda 2\slash{n}_\lambda U_{-\nu }\left( b_{-1}\right)
=\delta_{\lambda, \nu} U_\lambda \left( b_{-1}\right)
\label{stp11}
\end{equation}
we can fix the phases of the spinors $U_\lambda
\left(b_{-1}\right)$ and $U_\lambda \left( b_{1}\right)$.

By using the properties of lightlike vectors (\ref{stp3}) and
$\gamma$-matrix algebra it is possible to determine that
\begin{equation}
\frac{\slash{b}_{-1}}{2} U_{-\lambda
}\left(b_{1}\right)=U_{\lambda} \left( b_{-1}\right),
\label{stp12}
\end{equation}
\begin{equation}
\frac{\lambda}{2}\slash{n}_\lambda U_{\nu} \left(
b_1\right)=-\delta_{\lambda, \nu} U_{-\lambda }\left( b_1\right).
\label{stp13}
\end{equation}

The important property of basis spinors (\ref{stp8}), (\ref{stp9})
is the completeness relation
\begin{equation}
\frac{1}{2} \sum_{\lambda,A=-1}^{1} U_\lambda \left(
b_A\right) \overline{U}_{-\lambda} \left(b_{-A}\right)= I,
\label{stp14}
\end{equation}
which follows from Eqs.(\ref{stp8})-(\ref{stp10}). Thus, the
arbitrary bispinor can be decomposed in terms of basis spinors
$U_{\lambda}\left(b_{A} \right)$.

With the help of Eqs.(\ref{stp7}),(\ref{stp11})-(\ref{stp13}) we
can obtain that
\begin{equation}
\gamma^\mu U_\lambda \left(b_{-1}\right)= b_{-1}^{\mu}
U_{-\lambda} \left(b_1\right) + \lambda n_{\lambda}^{\mu}
U_{-\lambda} \left(b_{-1}\right), \label{stp15}
\end{equation}
\begin{equation}
\gamma^\mu U_\lambda \left(b_1\right)= b_1^{\mu} U_{-\lambda}
\left(b_{-1}\right) + \lambda n_{-\lambda}^{\mu} U_{-\lambda}
\left(b_1\right). \label{stp16}
\end{equation}
Eqs.(\ref{stp15}) and (\ref{stp16}) can be rewritten in a general
form
\begin{equation}
\gamma ^\mu U_\lambda \left( b_A\right) =b_A^\mu U_{-\lambda
}\left( b_{-A}\right) +\lambda n_{-A \ast\lambda }^\mu U_{-\lambda
}\left( b_A\right). \label{stp17}
\end{equation}

Another important property of basis spinors is that spinor
products are simple and are similar to scalar products of
isotropic tetrad vectors
\begin{equation}
\overline{U}_\lambda \left( b_C\right) U_{\rho}\left( b_A\right) =
2 \delta_{\lambda,-\rho} \delta_{C,-A},~~~~C,A=\pm 1
 ~\lambda,\rho=\pm 1.
\label{stp18}
\end{equation}
Eqs.(\ref{stp17}), (\ref{stp18}), and also
\begin{equation}
\label{stp19}
\omega_{\lambda}U_\rho \left(b_A\right)=\delta_{\lambda,
\rho} U_\rho \left(b_A\right)
\end{equation}
underlie the suggested method (MBS).

Let us consider some properties of construction
\begin{equation}
\Gamma _{C,A;\sigma
,\rho}^{\left\{\alpha,\beta,\ldots,\mu\right\}} \equiv
\overline{U}_\sigma \left(
b_C\right)\gamma_{\alpha}\gamma_{\beta}\ldots \gamma_{\mu} U_\rho
\left( b_A\right). \label{stp41b}
\end{equation}
By means of relations (\ref{stp17}) and (\ref{stp18}) it is easy
to calculate $\Gamma _{C,A; \sigma,
\rho}^{\left\{\alpha,\beta,\ldots,\mu\right\}}$ in terms of the
isotropic tetrad vectors. For example, the current-like
construction $\Gamma _{C,A;\sigma ,\rho }^{\left\{\mu\right\}}$
has the following form
\begin{equation}\label{stp40b}
\Gamma _{C,A;\sigma ,\rho }^{\left\{\mu\right\}}= 2 \delta_{\sigma
, \rho} \left(\delta_{C,A} b_{A}^{\mu}+\rho n_{A \ast \rho
}^{\mu}\delta_{-C,A} \right).
\end{equation}
Eq.(\ref{stp40b}) can be rewritten in matrix form
\begin{equation}\label{stp40bb}
\Gamma _{C,A;\sigma ,\rho }^{\left\{\mu\right\}}= 2
\delta_{\sigma, \rho}\left (
\begin{array}{cc}
b_{1}^{\mu} & \rho n_{-\rho}^{\mu}  \\
\rho n_{\rho}^{\mu} &  b_{-1}^{\mu}
  \end{array}
\right)_{CA}.
\end{equation}
With the help of completeness relation (\ref{stp14}) we can
receive recursion formula for $\Gamma_{-C,A;\sigma ,\rho}$
\begin{eqnarray}
&& \Gamma _{-C,A;\sigma ,\rho}^{\left\{\alpha,\beta,\ldots,\mu
\right\}}=1/2 \sum_{D=-1}^{1}\Gamma _{-C,D;\sigma, \sigma}^{
\left\{\alpha\right\}} \Gamma _{-D,A;-\sigma , \rho}^{
\left\{\beta,\ldots,\mu \right\}}
\nonumber\\
&& =1/4 \sum_{D,B=-1}^{1}\Gamma _{-C,D;\sigma, \sigma}^{
\left\{\alpha \right\}} \Gamma _{-D,B;-\sigma , -\sigma}^{
\left\{\beta \right\}} \Gamma _{-B,A;\sigma , \rho}^{
\left\{\ldots, \mu \right\}}.
\label{stp19e}
\end{eqnarray}

By means of the additional construction
\begin{equation}\label{stp19c}
  \Gamma _{C,A;\sigma ,\rho}^{n}=k_{1}^{\alpha}
k_{2}^{\beta}\ldots k_{n}^{\mu} \Gamma _{C,A;\sigma
,\rho}^{\left\{\alpha,\beta,\ldots,\mu \right\}},
\end{equation}
where $k_1,k_2,\ldots k_n$ are the arbitrary four-vectors,
recursion relation (\ref{stp19e}) can been rewritten in matrix
form
\begin{eqnarray}
&&  \Gamma _{-C,A;\sigma ,\rho}^{n}=2 \delta_{\sigma,
(-1)^{n-1}\ast\rho} \prod_{j=1}^{n} \left(
\begin{array}{cc}
(-1)^{j-1}\rho \left (n_{(-1)^{j}\ast\rho}\cdot k_{j} \right) & \left( b_{1} \cdot k_{j} \right) \\
 \left( b_{-1}\cdot k_{j} \right) & (-1)^{j-1}\rho \left (n_{(-1)^{j-1}\ast\rho}\cdot k_{j} \right)
  \end{array}
\right) \nonumber\\
&& =2 \delta_{\sigma, (-1)^{n-1}\ast\rho} \left(
\begin{array}{cc}
\rho \left (n_{-\rho}\cdot k_{1} \right) & \left( b_{1} \cdot k_{1} \right) \\
 \left( b_{-1}\cdot k_{1} \right) & \rho \left (n_{\rho}\cdot k_{1} \right)
  \end{array}
\right)\left(
\begin{array}{cc}
-\rho \left (n_{\rho}\cdot k_{2} \right) & \left( b_{1} \cdot k_{2} \right) \\
 \left( b_{-1}\cdot k_{2} \right) & -\rho \left (n_{-\rho}\cdot k_{2} \right)
  \end{array}
\right)\ldots
 ~~~. \label{stp19b}
\end{eqnarray}

Thus, the construction $\Gamma^{\alpha,\beta,\ldots, \mu}$ can be
easily calculated for a great number of $\gamma$-matrices with
the help of recursive relation (\ref{stp19b}). These calculations
are an important part of the method of basis spinors. The
algorithm of the method of basis spinors will be explained in
Sec.\ref{sec:level4}.

\section{Decomposition of the bispinors in terms of basis spinors}
\label{sec:level3}

The major part of my method is the calculation of decomposition
coefficients of an arbitrary bispinor on basis spinors
(\ref{stp8})-(\ref {stp9}). The opportunity of deriving these
coefficients is founded on that an arbitrary bispinor of a fermion
can be determined through the basis spinor
$U_\rho\left(b_{-1}\right)$ (or $U_\rho\left(b_1\right)$) with
the help of projection operators.

Let us consider massless fermions. An arbitrary bispinor
$U_\lambda \left(p\right)$ of momentum $p$ ($p^2=0, \left(p \cdot
b_{-1}\right) \not = 0$)  and helicity $\lambda$ is defined in
terms of basis spinor (see, for example, \cite{Kleiss})
\begin{equation}
U_\lambda \left(p\right) = \frac{\slash{p}}{\sqrt{2 \hskip 1pt
\left(p \cdot b_{-1}\right) }} U_{-\lambda }\left(b_{-1}\right).
\label{stp20}
\end{equation}

As it follows from completeness relation (\ref{stp14}), the
decomposition coefficients of arbitrary bispinors are the spinor
products
\begin{equation}\label{stp21}
D_{\lambda,\rho}\left(A;p\right)=\frac{1}{2} \overline{U}_\lambda
\left(b_A\right)U_\rho \left(p\right).
\end{equation}
Using Eqs.(\ref{stp17})-(\ref{stp20}) we obtain, that
\begin{equation}
D_{\lambda,\rho}\left(A;p\right)=
\frac{\delta_{\lambda,-\rho}}{\sqrt{2}}\left[
\delta_{A,-1}\sqrt{\left(p \cdot b_{-1}\right)}+
\delta_{A,1}\frac{\lambda \hskip 1pt \left(p \cdot
n_{\lambda}\right)}{\sqrt{\hskip 1pt \left(p \cdot
b_{-1}\right)}}\right]. \label{stp22}
\end{equation}

If $p=const~b_{-1} $, the decomposition has the most simple form,
as in this case
\begin{equation}\label{stp23}
U_\lambda \left(p\right) =\sqrt{const}~
U_{\lambda}\left(b_{-1}\right).
\end{equation}
For numerical calculations, as well as in the  case of using
spinor techniques, it is convenient to determine the coefficient
(\ref{stp22}) through the momentum components $p =\left
(p^0\right.$,$ p^x=p^0 \sin\theta_p\sin\varphi_p$, $p^y= p^0
 \sin\theta_p\cos\varphi_p$, $\left. p^z=p^0 \cos\theta_p\right)$
\begin{eqnarray}
&& D_{\lambda,\rho}\left(A;p\right)=
\frac{\delta_{\lambda,-\rho}}{\sqrt{2}}\left[ \delta_{A,-1}\sqrt{
p^{+}}- \delta_{A,1}\lambda \exp\left(i \lambda
\varphi_{p}\right)\sqrt{p^{-}}\right]
\nonumber\\
&&=\delta_{\lambda,-\rho}\sqrt{p^0}\left[ \delta_{A,-1}\cos
\frac{\theta_p}{2}- \delta_{A,1}\lambda \sin
\frac{\theta_p}{2}\exp\left(i \lambda \varphi_{p}\right) \right],
\label{stp24}
\end{eqnarray}
where
$$
p^{\pm}=p^0\pm p^z, ~~ p^x+i \lambda p^y=\sqrt{\left(p^x
\right)^2+\left(p^y \right)^2} \exp\left(i \lambda
\varphi_{p}\right).
$$

The decomposition of the bispinor of antifermion $V_\lambda
\left(p\right)$ is received from the ratio
\begin{equation}
V_\lambda \left(p\right) =U_{-\lambda }\left( p\right).
\label{stp25}
\end{equation}

Let us consider massive Dirac particles. The bispinors of the
massive fermion and anti-fermions with arbitrary polarization
vectors are determined by the basic spinor (see the Appendix in
\cite{Kleiss1}).
\begin{equation}
U_\lambda \left( p,s_p \right) =\frac{\tau _u^\lambda \left( p,s_p
\right) } {\sqrt{\left( b_{-1} \cdot \left( p+m_p
s_p\right)\right)}}U_{-\lambda }\left( b_{-1}\right) ,
\label{stp26}
\end{equation}
\begin{equation}
V_\lambda \left( p,s_p\right) = \frac{ \tau _v^\lambda
\left(p,s_p\right)} { \sqrt{\left( b_{-1} \cdot\left( p+m_p s_p
\right)\right)} } U_\lambda \left( b_{-1} \right), \label{stp27}
\end{equation}
where the projection operators $\tau _u^\lambda \left(
p,s_p\right), \tau _v^\lambda \left(p, s_p\right)$ are expressed
by the relations
\begin{equation} \tau _u^\lambda
\left( p,s_p\right) =\frac 1{2}\left( \slash{p}+m_p\right) \left(
1+\lambda \hskip 2pt \gamma _5 \slash{s}_p\right), \label{stp28}
\end{equation}
\begin{equation}
\tau _v^\lambda \left(p, s_p\right) =\frac 1{2}\left(
\slash{p}-m_p\right) \left(1+\lambda \gamma _5 \slash{s}_p\right).
\label{stp29}
\end{equation}
We obtain
\begin{eqnarray}
&&\slash{p} \hskip 2pt U_\lambda \left(p, s_p\right) = m_p \hskip
2pt U_{\lambda }\left(p, s_p\right), \hskip 7pt \slash{p} \hskip
2pt V_\lambda \left(p, s_p\right) = - m_p \hskip 2pt V_{\lambda
}\left(p, s_p\right) , \nonumber\\&& \gamma_5 \slash{s}_p \hskip
2pt U_\lambda \left(p, s_p\right) = \hskip 7pt \lambda \hskip 2pt
U_{\lambda }\left(p, s_p\right), \gamma_5\slash{s}_p \hskip 2pt
V_\lambda \left(p, s_p\right) = \lambda \hskip 2pt V_{\lambda
}\left(p, s_p\right) \label{stp30}
\end{eqnarray}
i.e. the bispinors $U_\lambda \left( p,s_p\right)$ and $V_\lambda
\left( p,s_p\right) $ satisfy Dirac equation and spin condition
for massive fermion and antifermion. The notation $s_p$ for
bispinors indicates that fermion with momentum $p$ has fixed
polarization vector $s_p$. We also found, that the bispinors of
fermions and antifermions,  Eqs.(\ref{stp26}),(\ref{stp27}), were
related by
\begin{equation}
V_\lambda \left( p,s_p\right)= -\lambda \gamma _5U_{-\lambda
}\left( p,s_p\right), \hskip 7pt \overline{V}_\lambda \left(
p,s_p\right) =\overline{U}_{-\lambda }\left( p,s_p\right)
\lambda \hskip 2pt \gamma_5. \label{stp31}
\end{equation}

After evaluations with the help of
Eqs.(\ref{stp26}),(\ref{stp27}) we receive, that the
decomposition coefficients for a massive fermion with momentum
$p$, an arbitrary polarization vector $s_p$ and mass $m_p $ can
be written as scalar products of tetrad and physical vectors
\begin{eqnarray}
&& D_{\lambda,\rho}\left(A;p,s_p\right)= \frac 1{\sqrt{2
\left(b_{-1} \cdot \xi _1^p \right)}}
\nonumber\\
&& \left[ \delta_{A,-1} \left\{ \delta _{\lambda,-\rho}
\left(b_{-1} \cdot \xi _1^p\right) -
\frac{\rho\delta_{\lambda,\rho}}{2 m_p}\left( \left(b_{-1} \cdot
\xi_1^p\right) \left(n_{-\rho} \cdot \xi _2^p\right)
+\left(n_{-\rho} \cdot \xi _1^p\right) \left( b_{-1} \cdot \xi
_2^p\right) \right) \right\} \right. +
\nonumber\\
&&\left. \delta_{A,1}\left\{- \rho \delta _{\lambda,-\rho}
\left(n_{-\rho} \cdot \xi _1^p\right)+ \frac{\delta _{\lambda,
\rho }}{2 m_p} \left(\left(b_{-1} \cdot \xi _1^p\right)\left(b_1
\cdot \xi _2^p\right)-\left(n_{-\rho} \cdot \xi _1^p\right)
\left(n_{\rho} \cdot \xi _2^p \right)\right) \right\} \right] ,
\label{stp32}
\end{eqnarray}
\begin{equation}
\xi _1^p =p+m_p s_p,~~~ \xi _2^p =p-m_p s_p . \label{stp33a}
\end{equation}

Let's determine decomposition coefficients for helicity and $KS$
polarization states of fermions, as they are the most-used in
calculations of matrix elements. The polarization vector of
$KS$-states is defined as follows
\cite{Kleiss,Andreev,Ballestrero}:
\begin{equation}
s_p = \frac{p}{m_p}-m_p \hskip 2pt \frac{b_{-1}}{\left(p \cdot
b_{-1}\right)} . \label{stp33}
\end{equation}
In this case, the massive fermion bispinor is related with basis
bispinor most simply \cite{Kleiss,Ballestrero}
\begin{equation}\label{stp34}
U_\lambda \left(p,KS\right) = \frac{\slash{p}+m_p}{\sqrt{2 \hskip
1pt \left(p \cdot b_{-1}\right) }} U_{-\lambda
}\left(b_{-1}\right)
\end{equation}
and the decomposition coefficients have a compact form
\begin{equation}\label{stp35}
D_{\lambda,\rho}\left(A;p,KS\right)=\frac{1}{\sqrt{2}}\left\{
\delta_{\lambda,-\rho}\left[ \delta_{A,-1}\sqrt{\left(p \cdot
b_{-1}\right) }+ \delta_{A,1}\frac{\lambda \hskip 1pt \left(p
\cdot n_{\lambda}\right)}{\sqrt{\left(p \cdot
b_{-1}\right)}}\right]+\delta_{\lambda,\rho}\delta_{A,1}\frac{m_p}
{\sqrt{\left(p \cdot b_{-1}\right)}}\right\}.
\end{equation}

Choosing
\begin{equation}
s_p=\frac{\left(p \cdot l_0\right) \hskip 2pt p -m^2_p \hskip 2pt
l_0} {m_p\sqrt{(p \cdot l_0)^2-m^2_p}} \label{stp36}
\end{equation}
we find that the polarization state of a fermion is its helicity
state.

For helicity states the expression for decomposition coefficients
in terms of scalar products of physical and isotropic tetrad
vectors  have more complex form, than for $KS$-states. But, if to
consider this expression through components of momentum
$p=\left(p^0,\left|\overrightarrow{p}\right|\sin\theta_p\sin\varphi_p\right.$,
$\left|\overrightarrow{p}\right|\sin\theta_p\cos\varphi_p$,
$\left. \left|\overrightarrow{p}\right|\cos\theta_p\right)$ we
obtain that the decomposition coefficients have a simple form
\begin{eqnarray}
&&D_{\lambda,\rho}\left(A;p,Hel\right)=\frac{1}{\sqrt{2}}\left[
\sqrt{p^0+\left| \overrightarrow{p}\right| }\left( \delta _{A,-1}
\cos\frac{\theta_p}{2} -\delta _{A,1} \lambda \exp \left( i
\lambda \varphi _p\right)\sin \frac{\theta_p}{2}\right) \delta
_{\lambda ,-\rho }\right. +
\nonumber \\
&&\left. \sqrt{p^0-\left| \overrightarrow{p}\right| }\left(
\delta_{A,1} \cos \frac{\theta_p}{2} +\delta _{A,-1} \lambda \exp
\left( -i \lambda \varphi _p\right)\sin \frac{\theta_p}{2} \right)
\delta _{\lambda ,\rho }\right]. \label{stp37}
\end{eqnarray}
It is clear, if $m_p=0$ Eq.(\ref{stp37}) turns into
Eq.(\ref{stp24}).

The analysis of decomposition coefficients for massive and
massless fermions of a spin $1/2 $ shows, that the decomposition
coefficient $D_{\lambda, \rho}$ for massless case becomes
diagonal on spin indices $\lambda, \rho$. This fact simplifies the
calculation of matrix elements with massless fermions.

The decomposition coefficients for an antifermion with  bispinor
(\ref{stp27}) can be easily obtained with the help of expression
(\ref{stp31}).

\section{Method of the Basis Spinors}
\label{sec:level4}

The amplitude of the Feynman diagram is the  product of the
expressions of the type
\begin{equation}
M_{\lambda _p,\lambda _k}\left( p,s_p;k,s_k\right)=
\overline{U}_{\lambda _p}\left( p,s_p\right) Q ~U_{\lambda
_k}\left(k,s_k\right).  \label{stp38}
\end{equation}

Equation (\ref{stp38}) corresponds to the fermion line with the
matrix operator $Q$, which is expressed as the combination of
$\gamma$-matrices and their contractions with four-vectors.

In the method based on the use of traces Eq.(\ref{stp38}) is
rewritten as follows
\begin{equation}
M_{\lambda _p, \lambda _k}\left(p,s_p; k,s_k\right) = Tr\left(
U_{\lambda_k}\left(k, s_k\right) \otimes \overline{U}_{\lambda
_p}\left(p,s_p\right) Q\right).  \label{stp38aa}
\end{equation}

The main problem of the trace approach is to get explicit form of
expression $U_{\lambda_k}\left(k,s_k\right) \otimes
\overline{U}_{\lambda _p}\left(p,s_p\right)$. A great number of
methods of matrix element calculation are directed to the
solution of this problem. In our case we can also obtain the
expression of the matrix $U_{\lambda_k}\left(k,s_k\right) \otimes
\overline{U}_{\lambda _p}\left(p,s_p\right)$ by means of Eqs.
(\ref{stp26}) and (\ref{stp27}) for various spin configurations
of fermions
\begin{equation}
U_\lambda \left(k, s_k\right) \otimes \overline{U}_\lambda \left(
p,s_p\right) =\frac{\tau _\lambda ^u \left(k, s_k\right) \omega
_{-\lambda } \slash{b}_{-1}\tau _\lambda ^u \left( p,s_p\right)
}{4\sqrt{b_{-1}\cdot \left(p+m_p s_p\right) }\sqrt{b_{-1}\cdot
\left(k+m_k s_k\right) }},  \label{stp38b}
\end{equation}
\begin{equation}
U_{-\lambda }\left(k, s_k\right) \otimes \overline{U}_\lambda
\left(p, s_p\right) =\frac{\lambda \tau _{-\lambda }^u\left(k,
s_k\right) \slash{n}_\lambda \omega _\lambda \slash{b}_{-1} \tau
_\lambda ^u\left(p, s_p\right) }{8 \sqrt{b_{-1} \cdot  \left(
p+m_p s_p\right) }\sqrt{b_{-1} \cdot\left( k+m_k s_k\right) }}.
\label{stp38c}
\end{equation}

Thus, to calculate (\ref{stp38}) it is necessary to evaluate the
relations:
\begin{equation}
M_{\lambda ,\lambda }\left(p,s_p;k,s_k\right) =\frac {Tr\left(
\tau _\lambda ^u\left(k,s_k\right) \omega _{-\lambda
}\slash{b}_{-1}\tau _\lambda ^u\left( p,s_p\right) Q\right)}
{4\sqrt{b_{-1} \cdot \left( p+m_p s_p\right) }\sqrt{b_{-1} \cdot
\left(k+ m_k s_k\right)}}, \label{stp38d} \end{equation}
\begin{equation}
M_{\lambda ,-\lambda }\left(p,s_p;k,s_k\right) =\frac {\lambda
Tr\left(\tau _{-\lambda }^u\left(k,s_k\right) \slash{n}_\lambda
\omega _\lambda \slash{b}_{-1}\tau _\lambda ^u \left( p,s_p\right)
Q\right)} {8\sqrt{ b_{-1}\cdot \left(p+m_p s_p\right)
}\sqrt{b_{-1} \cdot \left( k+m_k s_k\right)}}.  \label{stp38e}
\end{equation}

The main inconvenience of the trace method is the considerable
increase of terms with the increase of the number of
$\gamma$-matrices in the operator $Q$. In particular, the authors
of \cite{Ilyin} suggest to use $KS$ polarization states to
decrease the number of terms in $U_{\lambda_k}\left(k,s_k\right)
\otimes \overline{U}_{\lambda _p}\left(p,s_p\right)$. It is also
necessary to evaluate spin configurations of fermions separately.

In the paper I present the method of calculating expressions
(\ref{stp38}) without using the above scheme, i.e. without using
the traces of type (\ref{stp38d}), (\ref{stp38e}). This approach
can be realized by using the properties of basis spinors
(\ref{stp8}), (\ref{stp9}). That's why the method is called the
method of basis spinors (MBS).

The essence of the MBS is as follows. With the help of
completeness relation (\ref{stp14}) we decompose bispinors in
(\ref{stp38}). As a result the matrix element is
\begin{eqnarray}
&& M_{\lambda _p,\lambda _k}\left( p,s_p;k,s_k\right)= \frac{1}{4}
\sum_{\sigma ,\rho =-1}^1\sum_{A,C=-1}^1
\overline{U}_{\lambda_p}\left(
p,s_p\right)U_{\sigma}\left(b_{C}\right)
\nonumber\\
&& \left\{\overline{U}_{-\sigma}\left(b_{-C}\right) Q~
U_{\rho}\left( b_{A}\right)\right\}\overline{U}_{-\rho}\left(
b_{-A}\right) U_{\lambda _k}\left(k,s_k\right)
\nonumber\\
&& =\sum_{\sigma ,\rho =-1}^1\sum_{A,C=-1}^1 D_{\lambda _p,\sigma
}^{\dagger}\left( C;p,s_p\right) \Gamma _{-C,A;-\sigma ,\rho
}\left(Q \right) D_{-\rho ,\lambda _k}\left( -A;k,s_k\right),
\label{stp41ab}
\end{eqnarray}
where the coefficients $D_{\rho ,\lambda}$ are the decomposition
coefficients of the bispinors, and the construction $\Gamma$ is
defined by the relation
\begin{equation}
\Gamma _{C,A;\sigma ,\rho }\left(Q\right) \equiv
\overline{U}_\sigma \left( b_C\right) Q ~U_\rho \left( b_A\right).
\label{stp41bb}
\end{equation}
Equation (\ref{stp41ab}) can be rewritten in matrix form
\begin{eqnarray}
&&M_{\lambda _p,\lambda _k}\left( p,s_p;k,s_k\right)=
\sum_{\sigma,\rho=-1}^{1}
\nonumber  \\
&& \left(
\begin{array}{cc}
D_{\lambda _p, \sigma}^{\dagger}\left( -1;p,s_p\right)  &
D_{\lambda _p,\sigma}^{\dagger}\left( 1;p,s_p\right)
\end{array}
\right) \left(
\begin{array}{cc}
\Gamma _{1,-1;-\sigma, \rho } \left(Q\right)& \Gamma _{1,1;-\sigma, \rho}\left(Q\right)\\
\Gamma _{-1,-1;-\sigma, \rho }\left(Q\right)& \Gamma
_{-1,1;-\sigma, \rho}\left(Q\right)
\end{array}
\right) \left(
\begin{array}{c}
D_{-\rho ,\lambda _k}\left( 1;k,s_k\right)  \\
D_{-\rho ,\lambda _k}\left( -1;k,s_k\right)
\end{array}
\right) . \label{app5}
\end{eqnarray}
Thus in the MBS the problem of calculating (\ref{stp38}) involves
two steps:
\begin{enumerate}
  \item The calculation of the decomposition coefficients $D_{\lambda _p,\sigma
}\left(C;p,s_p\right)$ and $ D_{\rho ,\lambda _k}\left(
A;k,s_k\right)$
  \item The calculation of the value $\Gamma _{C,A;\sigma ,\rho }\left(
  Q\right)$ with further summing.
\end{enumerate}

The first part of the problem is solved in Sec.\ref{sec:level3}.
Evidently, such calculation are made only once and further on the
decomposition coefficients are used as ready-made functions.

By means of Eqs.(\ref{stp18}),(\ref{stp19}) and
(\ref{stp40bb})-(\ref{stp19b}) it is easy to calculate the value
$\Gamma\left(Q\right)$ as linear combinations of values $\Gamma
_{C,A;\sigma ,\rho}^{ \left\{\alpha,\beta,\ldots,\mu,\nu\right\}}$
(\ref{stp19b}). We can obtain Eq.(\ref{stp41bb}) in terms of the
scalar products of physical vectors included in $Q$ and  the
vectors of an isotropic tetrad (see, Sec.~\ref{sec:level2} and
Appendix~\ref{sec:level7}).

The further procedure of summing is largely simplified because of
Kronecker symbols occurring in calculating both decomposition
coefficients and value (\ref{stp41bb}).

In the case when operator $Q$ contains  non-contracted
Lorentz-indices, the final result of Eq.(\ref{stp38}) will make
up the corresponding tensor constructed from the vectors of an
isotropic tetrad.

As an example of the MBS let's consider the calculation of
Eq.(\ref{stp38}) for massless fermions
\begin{equation}
M_{\lambda _p,\lambda _k}\left(p,k;Q\right)\equiv
\overline{U}_{\lambda _p}\left(p\right) Q U_{\lambda _k}\left(
k\right) \label{stp39}
\end{equation}

Using coefficients of decomposition (\ref{stp22})  and summing we
find expression (\ref{stp39}) in general form (for arbitrary $Q$)
\begin{eqnarray}
M_{\lambda _p,\lambda _k}\left(p,k;Q\right)  &=&D_{\lambda
_p,-\lambda _p}^{\dagger}\left(1; p\right) \left(D_{-\lambda
_k,\lambda _k}\left(1; k\right) \Gamma _{-1,-1;\lambda _p,\lambda
_k}\left(Q\right)+D_{-\lambda _k,\lambda _k}\left(
-1; k\right) \Gamma _{-1,1;\lambda _p,\lambda _k}\left(Q\right)\right) +  \nonumber \\
&&D_{\lambda _p,-\lambda _p}^{\dagger}\left(-1; p\right) \left(
D_{-\lambda _k,\lambda _k}\left(1; k\right) \Gamma _{1,-1;\lambda
_p,\lambda _k}\left(Q\right)+D_{-\lambda _k,\lambda _k}\left(-1;
k\right) \Gamma
_{1,1;\lambda _p,\lambda _k}\left(Q\right)\right)
\nonumber \\
&=&\frac{-\lambda _p\left( p\cdot n_{\lambda _p}\right)
}{2\sqrt{\left( p\cdot b_{-1}\right) }}\left( \sqrt{\left( k\cdot
b_{-1}\right)} \Gamma _{-1,1;\lambda _p,\lambda
_k}\left(Q\right)-\frac{\lambda _k\left( k\cdot n_{-\lambda
_k}\right) }{\sqrt{\left( k\cdot b_{-1}\right) }}\Gamma
_{-1,-1;\lambda
_p,\lambda _k}\left(Q\right)\right) +  \nonumber \\
&&\frac{\sqrt{\left(p \cdot b_{-1}\right) } }{2} \left( \sqrt{
\left( k\cdot b_{-1}\right)} \Gamma _{1,1;\lambda _p,\lambda
_k}\left(Q\right)-\frac{\lambda _k\left( k\cdot n_{-\lambda
_k}\right) }{\sqrt{\left( k\cdot b_{-1}\right) }}\Gamma
_{1,-1;\lambda _p,\lambda _k}\left(Q\right)\right) . \label{app4}
\end{eqnarray}
For case $Q=I$ we have spinor product, i.e.
\begin{equation}\label{stp41sp}
\overline{U}_{\lambda _p}\left(p\right) U_{\lambda _k}\left(
k\right) =\lambda_{p}~ \delta_{\lambda_{p},-\lambda_{k}} \frac{
\left(p \cdot b_{-1}\right) \left(k \cdot n_{\lambda_{p}}\right)
-\left(k \cdot b_{-1}\right) \left(p \cdot n_{\lambda_{p}}
\right)}{\sqrt{\left( b_{-1}\cdot p\right)\left(b_{-1}\cdot
k\right)}}.
\end{equation}

For case $Q=\gamma^{\mu}$ with the help of (\ref{stp40b}) we find
expression (\ref{app4}) in terms of scalar products of physical
vectors $p,k$ and the vectors of an isotropic tetrad
\begin{eqnarray}
&&M_{\lambda _p,\lambda_k }\left(p,k;~\gamma^{\mu}\right) \equiv
J_{\lambda _p,\lambda_k }^{\mu}\left(p,k\right)=  \nonumber
\\
&&\delta _{\lambda _p,\lambda _k}\left( \left[ \sqrt{\left( p
 \cdot b_{-1}\right) \left( k \cdot b_{-1}\right) }\right] b_1^\mu +\left[
\frac{\left( p \cdot n_{\lambda _p}\right) \left( k \cdot
n_{-\lambda _p}\right) }{\sqrt{\left(p \cdot b_{-1}\right) \left(
k \cdot b_{-1}\right) }}\right] b_{-1}^\mu \right. -
\nonumber \\
&&\left. \left[ \left( p \cdot n_{\lambda _p}\right) \sqrt{\frac{
\left( k \cdot  b_{-1}\right)}{\left( p \cdot
b_{-1}\right)}}\right] n_{-\lambda _p}^\mu - \left[\left( k \cdot
n_{-\lambda _p}\right) \sqrt{ \frac{\left( p \cdot
b_{-1}\right)}{\left( k \cdot b_{-1}\right)}} \right] n_{\lambda
_p}^\mu \right). \label{stp41a}
\end{eqnarray}
For completeness we have included expressions (\ref{stp39}) with
$Q=\gamma^{\mu} \gamma^{\nu}$ and  $Q=\gamma^{\mu}
\gamma^{\nu}\gamma^{\alpha}$ in Appendix \ref{sec:level7}. With
the help of these relations we have an opportunity to gain larger
``building'' blocks of the Feynman diagrams and to use them as
universal functions.

The similar calculation for massive fermions does not differ and
leads to the appearance of additional terms in the vectors of an
isotropic tetrad. It is necessary to emphasize that the increase
of the number of $\gamma$-matrices in the operator $Q$ does not
lead to the avalanche increase of the number of terms as it is in
the trace methods. The matrix element in the suggested approach is
obtained for arbitrary spin configurations of fermions.

Let's briefly compare the method of basis spinors with spinor
technique of calculating the processes with massless fermions
(\cite{Kleiss}). We recall some details of the spinor techniques
with small modifications. Instead of  vectors $k_0=(1,1,0,0)$ and
$k_1=(0,0,1,0)$ used in Ref.(\cite{Kleiss}) we use  $b_{-1}$ and
$n_{\lambda}$ accordingly.

The CALCUL spinor techniques includes the following operations.
\begin{enumerate}
  \item An arbitrary massless spinor $U_\lambda \left(
p \right)$  is determined through the basic spinor by relation
(\ref{stp20}), i.e.
\begin{equation}
U_\lambda \left(p\right) = \frac{\slash{p}}{\sqrt{2 \hskip 1pt
\left(p \cdot b_{-1}\right) }} U_{-\lambda }\left(b_{-1}\right).
\label{stp20aa}
\end{equation}
\item The spinor Chisholm identity
\begin{equation}
\gamma ^\mu \left\{ \overline{U}_\lambda \left( p\right) \gamma
_\mu U_\lambda \left( k\right) \right\} =2 \left[ \hskip 1pt
U_\lambda \left( k\right) \overline{U}_\lambda \left( p\right) +
\hskip 1pt U_{-\lambda }\left( p\right) \overline{U}_{-\lambda
}\left(k\right)\right] \label{stp43}
\end{equation}
is used.
\item The four-vector $p$ with $p^2=0$ can be written as the sum of the
projection operators
\begin{equation}
\slash{p}=\sum_{\lambda=-1}^{1} U_\lambda \left( p\right)
\overline{U}_\lambda \left( p\right). \label{stp44}
\end{equation}
\item
The circular polarization vectors of massless boson
$\varepsilon^{\mu}_{\lambda}\left(k\right)$ with momentum $k$
($k^2=0$) is determined by
\begin{equation}
\label{stp45} \varepsilon^{\mu}_{\lambda}\left(k\right) \sim
\overline{U}_{\lambda}\left(q\right)\gamma^{\mu}
U_{\lambda}\left(k\right).
\end{equation}

With the help of Eqs.(\ref{stp43})-(\ref{stp45}) we can reduce
the amplitudes of processes to expressions involving only spinor
products of the type
\begin{equation}
s_{\lambda}\left(p,k\right) \equiv \overline{U}_\lambda
\left(p\right) U_{-\lambda} \left( k\right)=-s_{\lambda
}\left(k,p\right). \label{stp46}
\end{equation}
\end{enumerate}

Spinor product (\ref{stp46}) due to Eqs.(\ref{stp7}),(\ref{stp8})
and  (\ref{stp20aa}) is reduced to the calculation of trace
\cite{Kleiss}
\begin{eqnarray}
&&s_\lambda \left(p,k\right) = \frac{\lambda}{4}\frac{
Tr\left(\omega _{-\lambda }\slash{b}_{-1} \hskip 1pt \slash{p}
\hskip 1pt \slash{k} \hskip 1pt \slash{n}_\lambda \right)}
{\sqrt{\left(b_{-1} \cdot p\right)}\sqrt{\left(b_{-1} \cdot
k\right)}}
\nonumber\\
&& =\frac{ \lambda \left[ \left(p \cdot b_{-1}\right) \left(k
\cdot n_\lambda\right) -\left(k \cdot  b_{-1}\right) \left(p \cdot
n_\lambda \right) \right] - i\epsilon \left(b_{-1}\hskip 1pt,
n_\lambda \hskip 1pt, p \hskip 1pt, k \right)}{2\sqrt{\left(
b_{-1}\cdot p\right)\left(b_{-1}\cdot k\right)}} \label{stp47}
\end{eqnarray}
with
$\epsilon\left(p,r,k,q\right)=\epsilon^{\alpha\beta\rho\mu}p_{\alpha}
r_{\beta} k_{\rho} q_{\mu}$.

Using properties of isotropic tetrad vectors Eq.(\ref{stp47}) can
be rewritten
\begin{equation}\label{stp48}
s_\lambda \left(p,k\right) =\frac{\lambda \left[\left(p \cdot
b_{-1}\right) \left(k \cdot n_\lambda\right) -\left(k \cdot
b_{-1}\right) \hskip 2pt \left(p \cdot n_\lambda \right) \right]}
{\sqrt{\left( b_{-1}\cdot p\right)\left(b_{-1}\cdot k\right)}}.
\end{equation}

The comparative analysis of algorithms of matrix elements
reduction (calculation) shows that the spinor techniques and the
MBS differ from each other essentially. Neither Chisholm identity
(\ref{stp43}) nor additional constructions (\ref{stp44}),
(\ref{stp45}) are used in the MBS. The method of basis spinors
does not require a special procedure for constructing
polarization vectors of massive gauge bosons (see, for example
\cite{Kleiss}), as all four-vectors in the method are "processed"
similarly. Besides, it is necessary to mark that if in
Eq.(\ref{stp43}) the vector $p$ or $k$ coincides with $b_1$ or
$b_{-1}$ you should be very accurate using spinor Chisholm
identities, because they alter (see \cite{Jadach}).

The MBS can also be supplemented by constructing the polarization
vector of massless and massive bosons in the basis of an
isotropic tetrad (for photons, see \cite{Borodulin}). The final
expression for (\ref{stp39}) will contain only scalar products of
four-vectors of particles and the isotropic tetrad.

It is always possible to construct the basis of an isotropic
tetrad (\ref{stp3}) by means of physical vectors for particular
reaction. The procedure enables to obtain the expression of the
Feynman amplitude in terms of the scalar products of the
four-momenta and their contraction with Levi-Civita tensor. Thus,
we can get the matrix element in explicit Lorentz-covariant form.
Besides, the successful construction of the basis can lead to a
considerable decrease of the number of terms.

\section{Applications}
\label{sec:level5}

Let us consider a series of examples as an illustration and a
test of the method of basis spinors. We shall take
electron-positron reactions as there are enough examples of
analytical calculations of the matrix element for them. For
simplification we shall consider initial fermions as massless.

\begin{figure}[h t b p]
\begin{center}
\resizebox{0.6\textwidth}{!}{
\includegraphics{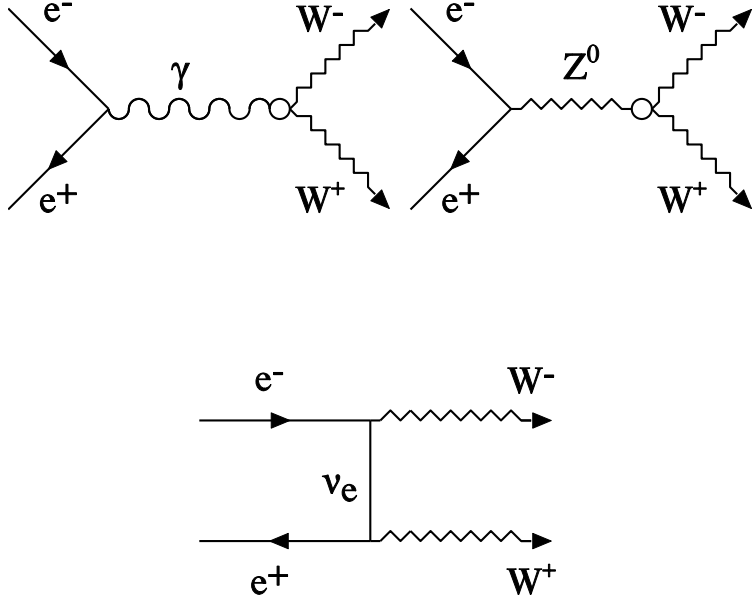}}
\end{center}
\vspace{-15mm} \caption{Feynman diagrams for the process $e^-
e^+\to W^- W^+$.} \label{eeww}
\end{figure}

As an example of our method we shall calculate the amplitudes of
the reaction
\begin{equation}
\label{stp49}
e^-\left(p_1,\lambda_1\right)+e^+\left(p_2,\lambda_2\right)\to
W^-\left(k_1, \alpha\right)+W^+\left(k_2,\beta \right).
\end{equation}
The Feynman diagrams are shown in Fig.\ref{eeww}.
The amplitude of the process can be written
\begin{eqnarray}
&&M_{e^{+}e^{-}\rightarrow W^{+}W^{-}} =M_{\gamma Z} +M_\nu ,
\label{stp50}
\\
&&M_{\gamma Z} = \frac{4\pi\alpha}{P^2} \left[
\overline{U}_{\lambda _2}\left( p_2\right) \gamma _\mu U_{\lambda
_1}\left( p_1\right)- \frac{P^2}{\left( P^2-m_Z^2\right) 2\sin
^2\theta _W} \right.\times \nonumber \\&& \left.
\overline{V}_{\lambda _2}\left( p_2\right) \gamma _\mu \left(
g_V^e-g_A^e\gamma _5\right) U_{\lambda _1}\left( p_1\right)
\right] \Gamma ^{\mu \alpha \beta }\left( P,k_1,k_2\right)
\varepsilon _\alpha \left( k_1\right) \varepsilon _\beta \left(
k_2\right),
\label{stp50a} \\
&&M_\nu =\frac{\pi \alpha }{2 Q^2\sin ^2\theta
_W}\overline{V}_{\lambda _2}(p_2)\hat{\varepsilon }\left(
k_2\right)\left( 1-\gamma _5\right) \hat{Q}\hat{ \varepsilon
}\left( k_1\right) \left( 1-\gamma _5\right) U_{\lambda _1}\left(
p_1\right) \label{stp50b},
\end{eqnarray}
where $g_A^e$, $g_V^e$ are axial and vector electron couplings
accordingly. The tensor $\Gamma ^{\mu \alpha \beta }\left(
P,k_1,k_2\right)$ is determined as
\begin{equation}
\Gamma ^{\mu \alpha \beta }\left( P,k_1,k_2\right) =g^{\alpha
\beta }\left( k_1-k_2\right) ^\mu +2\left( P^\alpha g^{\mu \beta
}-P^\beta g^{\mu \alpha }\right).   \label{stp51}
\end{equation}
In Eqs.(\ref{stp50a}) and (\ref{stp50b}) we use the following
designations: $\alpha$ and $\theta_W$ are the fine-structure
constant and weak mixing angle respectively, the vectors
$\varepsilon _\alpha \left( k_1\right)$, $\varepsilon _\beta
\left(k_2\right)$ are the polarization vectors of $W$-bosons and
the momenta $P,Q$ are $P=p_1+p_2$ $Q=p_1-k_1$.

Let us consider reaction (\ref{stp49}) in the center of momentum
system $e^+e^-$. Then vectors of isotropic tetrad (\ref{stp3})
can be expressed by means of physical vectors $p_1,p_2,k_1$
\begin{eqnarray}
&&b_1=\frac{\sqrt{2} p_1}{\sqrt{\left(p_1 \cdot p_2\right)}},~~~
b_{-1}=\frac{\sqrt{2} p_2}{\sqrt{\left(p_1 \cdot p_2\right)}},
\nonumber\\
&&n_\lambda=\frac{b_{-1}\left(k_1 \cdot b_1\right)+b_1\left(k_1
\cdot b_{-1}\right)-2 k_1+i\lambda\left[b_1, b_{-1}, k_1\right]}{2
\sqrt{\left(b_{-1} \cdot k_1\right) \left(b_1 \cdot k_1\right)}}
\label{stp52}
\end{eqnarray}
with
$\left[p,r,k\right]^{\mu}=\epsilon^{\alpha\beta\rho\mu}p_{\alpha}
r_{\beta} k_{\rho}$.

In this case it is easy to find with the help of the MBS that
\begin{equation}
\overline{V}_{\lambda _2}\left( p_2\right) \gamma ^\mu U_{\lambda
_1}\left( p_1\right) =\lambda _1 \delta _{-\lambda _2,\lambda
_1}\sqrt{2 \left( p_1 \cdot p_2\right)} n_{\lambda _1}^\mu .
\label{stp53}
\end{equation}
Using Eq.(\ref{stp53}) we obtain the matrix elements with
$\gamma$ and $Z^0$-boson exchanges in terms of scalar products of
the physical vectors and the isotropic tetrad vectors:
\begin{eqnarray}
 &&M_{\gamma Z}=
4\pi \alpha \frac{\lambda _1 \delta _{-\lambda _2,\lambda
_1}}{\sqrt{2 \left( p_1 \cdot p_2\right)}} \left(1
-\chi\left(P^2\right) \frac{g_e^{-\lambda_1}}{2\sin^{2}\theta _W
} \right)
\nonumber \\
&&\times \left( n_{\lambda _1}\right) _\mu \Gamma ^{\mu \alpha
\beta }\left( P,k_1,k_2\right) \varepsilon _\alpha \left(
k_1\right) \varepsilon _\beta \left( k_2\right)
\nonumber \\
&=&4\pi \alpha \frac{\lambda _1 \delta _{-\lambda _2,\lambda
_1}}{\sqrt{2 \left( p_1 \cdot p_2\right)}} \left(1
-\chi\left(P^2\right) \frac{g_e^{-\lambda_1}}{2\sin^{2}\theta _W }
\right)\left[ \left( \varepsilon \left( k_1\right) \cdot
\varepsilon \left( k_2\right) \right)
\left( \left(k_1-k_2\right) \cdot n_{\lambda _1}\right)\right. +  \nonumber \\
&&\left. 2\left(\left(P \cdot \varepsilon _\alpha \left(k_1\right)
\right) \left( \varepsilon \left(k_2\right)\cdot n_{\lambda
_1}\right) -\left( P \cdot \varepsilon \left( k_2\right) \right)
\left( \varepsilon \left(k_1\right)\cdot n_{\lambda _1}\right)
\right)\right], \label{stp54}
\end{eqnarray}
where we use the following notations:
$\chi\left(P^2\right)=P^2/\left(P^2-m_Z^2\right)$,
$g_e^{\lambda}=\left( g_V^e+\lambda g_A^e\right)$. It should be
noted that similar calculation by means of the spinor techniques
of the CALCUL group requires much more efforts.

Let us consider the calculation of the diagram with neutrino
exchange as an example of using building blocks, which were
calculated before by the method of basis spinors. Amplitude
(\ref{stp50b}) can be rewritten in the form
\begin{equation}
\label{stp63a} M_{\nu} =\frac{\pi \alpha
\left(1-\lambda_1\right)}{Q^2\sin ^2\theta _W}
\varepsilon_{\mu}\left(k_2\right) Q_{\nu}
\varepsilon_{\beta}\left(k_1\right) \overline{U}_{-\lambda
_2}\left(p_2\right)\gamma^{\mu} \gamma ^{\nu} \gamma ^{\beta}
U_{\lambda_1}\left(p_1\right).
\end{equation}
Using Eq.(\ref{app2}) from Appendix~\ref{sec:level7} and the
definition of isotropic tetrad vectors (\ref{stp52}) we obtain
\begin{eqnarray}
&&M_\nu  =\delta_{-\lambda_2,\lambda_1}\sqrt{2\left( p_1\cdot
p_2\right) }\frac{\pi \alpha \left(\lambda_1-1\right)}{Q^2\sin
^2\theta _W}\times
\nonumber \\
&& \left[ \left( \varepsilon \left( k_2\right) \cdot b_{-1}\right)
\left( \left( \varepsilon \left( k_1\right) \cdot b_1\right)
\left( n_{-\lambda _1}\cdot Q\right) -\left( \varepsilon \left(
k_1\right) \cdot n_{-\lambda _1}\right) \left( b_1\cdot
Q\right) \right) \right. +\nonumber \\
&&\left. \left( \varepsilon \left( k_2\right) \cdot n_{-\lambda
_1}\right) \left( \left( \varepsilon \left( k_1\right) \cdot
n_{-\lambda _1}\right) \left( n_{\lambda _1}\cdot Q\right) -\left(
\varepsilon \left( k_1\right) \cdot b_1\right) \left( b_{-1}\cdot
Q\right) \right) \right] \label{stp54n}.
\end{eqnarray}

To get an explicit Lorentz-covariant form of amplitude
(\ref{stp50}) it is necessary to substitute the vectors of
isotropic tetrad in relations (\ref{stp54}),(\ref{stp54n}) by
formulas (\ref{stp52}).

Taking into account the kinematics of the process the components
of the vectors in the center of momentum system $e^+e^-$ are
determined by the relations:
\begin{eqnarray}
&&p_1=\frac{\sqrt{s}}{2}(1,0,0,1),~~ p_2=\frac{\sqrt{s}}{2}
(1,0,0,-1),
\nonumber \\
&& k_1=\frac{\sqrt{s}}{2} (1, \beta_W \hskip 1pt \sin\hskip 1pt
\theta, 0, \beta_W \hskip 1pt \cos \hskip 1pt \theta ),
k_2=\frac{\sqrt{s}}{2} (1,-\beta_W \hskip 1pt \sin\hskip 1pt
\theta,0, -\beta_W \hskip 1pt \cos \hskip 1pt \theta),
\nonumber\\
&& \varepsilon ^{\mu}_{T} \left(k_1\right) =
\frac{1}{\sqrt{2}}\left( 0,\cos\theta ,\nu_1 i,-\sin \hskip 1pt
\theta \right) ,\varepsilon ^{\mu}_{T} \left(k_2\right) =
\frac{1}{\sqrt{2}}\left( 0,\cos\theta ,-\nu_2 i,-\sin \hskip 1pt
\theta \right) ,
\nonumber\\
&& \varepsilon ^{\mu}_{L} \left(k_1\right) = \gamma_W \left(
\beta_W,\sin\hskip 1pt \theta, 0,\cos \hskip 1pt \theta\right)
,\varepsilon ^{\mu}_{L} \left(k_2\right) = \gamma_W \left(
\beta_W,-\sin\hskip 1pt \theta, 0, -\cos \hskip 1pt \theta\right),
\label{stp55}
\end{eqnarray}
where $s=(p_1+p_2)^2$, $\beta_W=\sqrt{1-4 m_W^2/s}$,
$\gamma_W=\sqrt{s}/\left(2 m_W\right)$, ~$\nu_1,\nu_2=\pm 1$ and
the angle $\theta$ is the scattering angle of $W^-$-boson in the
center of momentum system . For longitudinally polarized
$W$-bosons ($\varepsilon \left(k_{1,2}\right)\equiv\varepsilon
_{L} \left(k_{1,2}\right)$) after a series transformations it is
not difficult to obtain
\begin{eqnarray}
&& M_{\gamma,Z}^{LL}=4\pi \alpha  \lambda _1 \delta _{-\lambda
_2,\lambda _1} \left( 1- \chi\left(s\right)
\frac{g_e^{-\lambda_1}}{2\sin^{2}\theta _W } \right)\beta_W \left(
2\gamma_W^2+1\right) \sin\theta, \label{stp56ll}
\\
&& M_{\nu}^{LL}= \frac{2 \pi \alpha \delta _{-\lambda _2,\lambda
_1}}{\beta_{W} \sin^{2} \theta _{W}}
\left(1-\lambda_1\right)\left( \gamma_W^2
-\frac{1}{\gamma_W^2\left(1+\beta_{W}^2-2 \beta_{W}\cos \theta
\right)} \right)\sin\theta .\label{stp56}
\end{eqnarray}
Eqs.(\ref{stp56ll}) and (\ref{stp56}) coincide with the matrix
elements of this process, which are presented in \cite{Hagiwara1}.

\begin{figure}[p h t b]
\begin{center}
\resizebox{0.5\textwidth}{!}{
\includegraphics{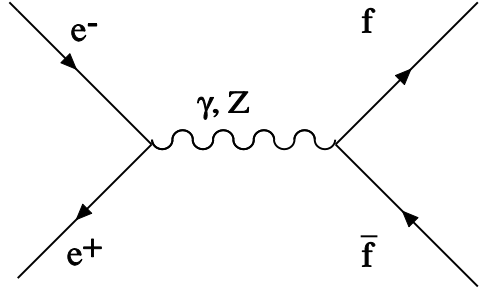}}
\end{center}
\vspace{-25mm}
\caption{Feynman diagrams for the process $e^-
e^+\to f \bar f$.} \label{eeff}
\end{figure}

We shall consider the process with massive fermions $f, \bar f$
\begin{equation}\label{stp57}
e^-\left(p_1,\lambda_1\right)+e^+\left(p_2,\lambda_2\right)\to
f\left(k_1,\nu_1\right)+\bar f\left(k_2,\nu_2\right)~~ (f\not = e)
\end{equation}
as the following test of the MBS. The Feynman diagrams for this
process are shown in Fig.\ref{eeff}.
The amplitude of the process can be written as
\begin{equation}
M_{e^+e^-\to f \bar f}\left (\lambda_1,\lambda_2;\nu_1,\nu_2
\right)= 4\pi\alpha/s \left[ M_{\gamma} \left
(\lambda_1,\lambda_2;\nu_1,\nu_2 \right)+ M_{Z^0} \left
(\lambda_1,\lambda_2;\nu_1,\nu_2 \right) \right], \label{stp58}
\end{equation}
where
\begin{equation}
M_{\gamma} \left ( \lambda_1,\lambda_2;\nu_1,\nu_2 \right) =Q_f
\overline{V}_{\lambda_{2} } \left(p_2 \right) \gamma_{\mu}
U_{\lambda_{1} } \left(p_1\right) \overline{U}_{\nu_{1} }
\left(k_1\right) \gamma^{\mu} V_{\nu_{2} } \left( k_2 \right),
\label{stp59}
\end{equation}
\begin{eqnarray}
&&M_{Z^0} \left ( \lambda_1,\lambda_2;\nu_1,\nu_2 \right)= R_Z
\left ( g^{\mu\nu}-P^\mu P^\nu/m_Z^2  \right ) \nonumber\\&&
\overline{V}_{ \lambda_{2} } \left( p_2 \right) \gamma_{\nu}
\left( g_V^e-g_A^e \gamma_5 \right ) U_{\lambda_{1}} \left(p_1
\right) \overline{U}_{\nu_{1}} \left( k_1 \right) \gamma_{\mu}
\left( g_V^f-g_A^f \gamma _5 \right ) V_{\nu_{2}} \left( k_2
\right)  \label{stp60}
\end{eqnarray}
with $R_Z=\left(G_F m_Z^2 s \right)/ \left(2 \sqrt{2}\pi \alpha
\left(s-m_Z^2 \right) \right)$. The values $g_V^f, g_A^f$ are
fermion coupling constants, $G_F$ is Fermi constant and $Q_f$ is
fermion charge $f$ in units $e$.

We shall consider the amplitudes of process (\ref{stp57}) for the
helicity and $KS$ polarization states of final fermions. Using
(\ref{stp53}) and the decomposition coefficients for $KS$
polarization states (\ref{stp35}) it is easy to obtain the
expressions both in terms of  scalar products and through
momentum components respectively by means of the MBS
\begin{eqnarray}
&&M_{e^+e^-\to f \bar f}^{KS} \left (
\lambda_1,\lambda_2;\nu_1,\nu_2 \right) =\frac{-8\pi \alpha
}{s}\frac{ \sqrt{s}\delta _{\lambda _1,-\lambda _2}}{\sqrt{\left(
b_{-1} \cdot k_1\right) \left( b_1 \cdot k_2\right) }}\left[
\delta _{\lambda _1,\nu _1}\left( b_{-1}\cdot k_1\right) \right.
\nonumber \\
&& \left( \lambda _2 Q_f-R_Z g_e^{-\lambda _1}g_f^{-\nu
_1}\right)\left( \left( n_{\nu _2}\cdot k_2\right) \delta _{\nu
_1,-\nu_2}+m_f \nu _2\delta _{\nu _1,\nu _2}\right) +\nu _2
\left(b_{-1}\cdot k_2\right) \nonumber \\&&
 \left(\nu_1\delta _{\nu
_1,-\nu _2}\delta _{-\lambda _1,\nu _1}\left( n_{\nu _1}\cdot
k_1\right) \left( -Q_f \lambda _2+R_Z g_e^{-\lambda _1}g_f^{-\nu
_1}\right)\right.
\nonumber \\
&& \left. \left. + m_f\left( Q_f\lambda _2-R_Z g_e^{-\lambda
_1}g_f^{\nu _1}\right) \right) \right], \label{stp61}
\end{eqnarray}
\begin{eqnarray}
&&M_{e^+e^-\to f\bar f}^{KS} \left (
\lambda_1,\lambda_2;\nu_1,\nu_2 \right) =\frac{ 4\pi \alpha \delta
_{\lambda _1,-\lambda _2}}{\sqrt{1-\beta _f^2\cos ^2\theta
}}\left[ \delta _{\nu _1,-\nu _2}\beta _f\left( \lambda _1Q_f+R_Z
g_e^{-\lambda _1}g_f^{-\nu _1}\right)  \right.
\nonumber \\
&&\sin \theta \left\{ \delta _{\lambda _1,\nu _1}\left( 1+\beta
_f\cos \theta \right) -\delta _{-\lambda _1,\nu _1}\left( 1-\beta
_f\cos \theta \right) \right\} +\delta _{\nu _1,\nu _2}\delta
_{\nu _1,\lambda _1}\nu _1\frac{2 m_f}{\sqrt{s} }
 \nonumber \\ && \left. \left\{ 2Q_f\lambda _1+R_Z g_e^{-\lambda _1}\left[ g_f^{-\nu
_1}\left( 1+\beta _f\cos \theta \right) +g_f^{\nu _1}\left(
1-\beta _f\cos \theta \right) \right] \right\} \right].
\label{stp62}
\end{eqnarray}

Using the MBS it is simple obtain the helicity amplitudes with the
help of the coefficients of decomposition (\ref{stp37}):
\begin{eqnarray}
&&M_{e^{+}e^{-}\to f\bar f}^{Hel}\left( \lambda _1,\lambda _2,\nu
_1,\nu _2\right) =4\pi \alpha \delta _{\lambda _1,-\lambda
_2}\left( \delta _{\nu _1,-\nu _2}\left\{ \left( \delta _{\lambda
_1,\nu _1}\left( 1+\cos \theta \right) -\delta _{\lambda _1,-\nu
_1}\left( 1-\cos \theta \right) \right)
\right. \right.   \nonumber \\
&&\left. \left[ 2Q_f+R_Z g_e^{-\lambda _1}\left( g_f^{\nu
_1}\left( 1-\beta _f\right) +g_f^{-\nu _1}\left( 1+\beta _f\right)
\right) \right] \right\} +
\nonumber \\
&&\left. \delta _{\nu _1,\nu _2}\left\{ \frac{\nu _1
m_f}{\sqrt{s}}\sin \theta \left[ 2 Q_f+R_Z g_e^{-\lambda _1 }
\left(g_f^{\nu _1}+g_f^{-\nu _1}\right) \right] \right\} \right) .
\label{stp62h}
\end{eqnarray}
The helicity amplitude (\ref{stp62h}) coincides with the
amplitude, which was obtained in Ref.\cite{Vega} up to the phase
factor.

It is easy to verify that the unpolarized cross sections, which
can be obtained from the $KS$ (\ref{stp62}) and helicity
amplitudes (\ref{stp62h}) coincide with the known expression
(see, for example, \cite{Borodulin}).

It should be noted, that $\delta$-symbols which occur in
(\ref{stp61}) and (\ref{stp62}),(\ref{stp62h}) are not
artificially introduced. The Kronecker symbols occur as the
consequence of the used method of calculations. That's why the
method gives the analytical expression of the matrix element for
all spin configurations of fermions at the same time. The
condition distinguishes the method from  the trace method where
the spin configurations of the fermions should be calculated
separately. Besides, it is convenient to use chiral (left and
right) coupling constants $(g_{f}^{\pm 1} \equiv g_{V}^{f}\pm
g_{A}^{f})$ in the MBS, because they occur during calculation
(\ref{stp41bb}).

\begin{figure}[h t b]
\begin{center}
\resizebox{0.6\textwidth}{!}{
\includegraphics{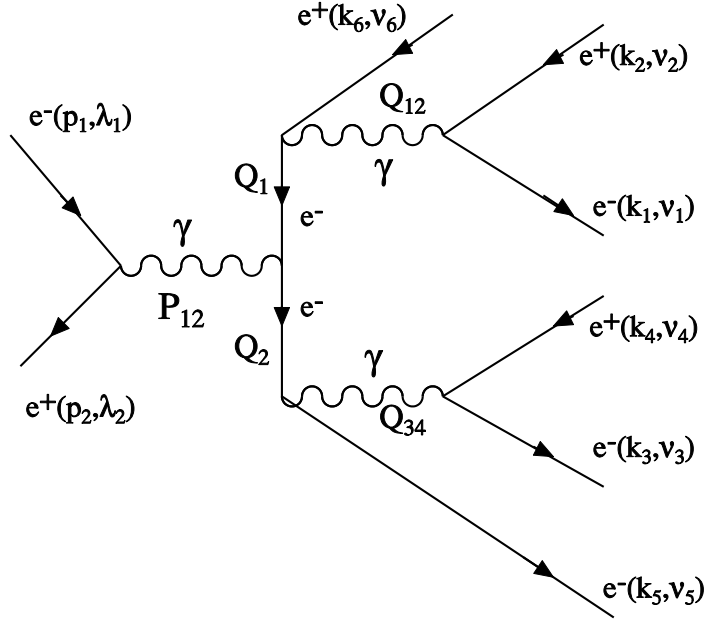}}
\vspace{-10mm}
\end{center}
\caption{\label{ee6f} One of the possible Feynman diagrams of
$e^+e^- \to e^+e^- e^+e^- e^+e^- $.}
\end{figure}

To avoid the impression that the method can calculate only simple
binary processes let's consider  the calculation of one of the
possible diagrams (see fig.\ref{ee6f}) for reaction $e^+e^- \to
e^+e^- e^+e^- e^+e^- $ with massless fermions as an illustration.
Using definition (\ref{stp41a}) the matrix element of given
Feynman diagram can be written as
\begin{eqnarray}
&&M_{e^{+}e^{-}\to e^{+}e^{-}e^{+}e^{-}e^{+}e^{-}}=\frac{e^6}{
P_{12}^2 Q_{12}^2 Q_{34}^2 Q_1^2 Q_2^2} J_{-\lambda _2,\lambda
_1}^\beta \left( p_2,p_1\right) J_{\nu _1,-\nu _2}^\rho \left(
k_1,k_2\right)
\nonumber \\
&&J_{\nu _3,-\nu _4}^\sigma \left( k_3,k_4\right)
\overline{U}_{\nu _5}\left( k_5\right) \gamma _\sigma
\hat{Q}_2\gamma _\beta \hat{Q} _1\gamma _\rho V_{\nu _6}\left(
k_6\right). \label{stp63}
\end{eqnarray}
Here we use the following notations for four-vectors
\begin{eqnarray}
&&Q_1 =-\left( k_1+k_2+k_6\right) ,Q_2 =k_3+k_4+k_5,
\nonumber \\
&&Q_{34}=k_3+k_4, Q_{12}=k_1+k_2, P_{12}=p_1+p_2. \label{stp64}
\end{eqnarray}
Using the MBS matrix element (\ref{stp63}) is reduced to compact
form
\begin{eqnarray}
&&M_{e^{+}e^{-}\to e^{+}e^{-}e^{+}e^{-}e^{+}e^{-}}=\frac{\delta
_{\nu _1,-\nu _2}\delta _{\nu _3,-\nu _4}\delta _{\nu _5,-\nu
_6}\delta _{\lambda _1,-\lambda _2}}{\sqrt{\left( k_5 \cdot
b_{-1}\right) \left(k_6 \cdot b_{-1}\right) }}\frac{\sqrt{s}
\lambda _1 e^6}{P_{12}^2 Q_{12}^2 Q_{34}^2 Q_1^2 Q_2^2}
\nonumber \\
&&\ \left( \delta _{\lambda _1,-\nu _5}\left[ \left( Q_1 \cdot
b_1\right) \left( \left( k_6  \cdot b_{-1}\right) \left( j_2 \cdot
n_{-\nu _5}\right) -\left( j_2 \cdot b_{-1}\right) \left( k_6
\cdot n_{-\nu _5}\right) \right) \right. \right. +
\nonumber \\
&&\ \left. \left(Q_1 \cdot n_{-\nu _5}\right) \left( \left(j_2
\cdot n_{\nu _5}\right) \left(k_6 \cdot  n_{-\nu _5}\right)
-\left( k_6 \cdot b_{-1}\right) \left(j_2 \cdot b_1\right)
\right) \right] \times  \nonumber \\
&&\ \left[ \left(Q_2 \cdot b_{-1}\right) \left( \left(k_5 \cdot
n_{\nu _5}\right) \left(j_3 \cdot n_{-\nu _5}\right) -\left(k_5
\cdot b_{-1}\right) \left(j_3 \cdot b_1\right) \right)
\right. +  \nonumber \\
&&\ \left. \left(Q_2 \cdot n_{-\nu _5}\right) \left( \left(k_5
\cdot b_{-1}\right) \left(j_3 \cdot n_{\nu _5}\right) -\left(j_3
\cdot b_{-1}\right) \left(k_5 \cdot n_{\nu
_5}\right) \right) \right] -  \nonumber \\
&&\ \delta _{\lambda _1,\nu _5}\left[ \left(k_6 \cdot n_{-\nu
_5}\right) \left( \left(j_2 \cdot b_{-1}\right) \left(Q_1 \cdot
n_{\nu _5}\right) -\left(Q_1 \cdot b_{-1}\right)
\left(j_2 \cdot n_{\nu _5}\right) \right) \right. +  \nonumber \\
&&\ \left. \left(k_6 \cdot b_{-1}\right) \left( \left(j_2 \cdot
b_1\right) \left(Q_1 \cdot b_{-1}\right) -\left(Q_1 \cdot n_{\nu
_5}\right) \left(j_2 \cdot n_{-\nu _5}\right)
\right) \right] \times  \nonumber \\
&&\ \left[ \left(Q_2 \cdot b_1\right) \left( \left(k_5 \cdot
b_{-1}\right) \left(j_3 \cdot n_{\nu _5}\right) -\left(j_3 \cdot
b_{-1}\right) \left(k_5 \cdot n_{\nu _5}\right)
\right) \right. +  \nonumber \\
&&\ \left. \left. \left(Q_2 \cdot n_{\nu _5}\right) \left( \left(
j_3 \cdot n_{-\nu _5}\right) \left(k_5 \cdot n_{\nu _5}\right)
-\left(k_5 \cdot b_{-1}\right) \left(j_3 \cdot b_1\right) \right)
\right] \right) , \label{e6c}
\end{eqnarray}
where
\begin{equation}\label{stp65a}
j_2^\mu \equiv J_{\nu _1,-\nu _2}^\mu \left(k_1,k_2\right),
~~j_3^\mu \equiv J_{\nu _3,-\nu _4}^\mu \left(k_3,k_4\right).
\end{equation}
The scalar products of the vectors $j_2,j_3$ with the vectors of
an isotropic tetrad are easily obtained with the help of
(\ref{stp41a}).
\section{Concluding Remarks}
\label{sec:level6}

The suggested computational method of matrix elements from the
methodological point of view (but not from the point of view of a
method of calculations) is close to the methods offered in
\cite{Hagiwara}, \cite{Ballestrero} for helicity and $KS$-fermion
states. In contrast to them the method of the basis spinors has a
simpler algorithm and does require the explicit form of
$\gamma$-matrices and bispinors. It can also be used for arbitrary
fermion polarization.

The calculation of a matrix element in the MBS is simplified
through the incorporation of a complete set of massless basis
spinors that makes many evaluations trivial. And the main
"laborious" operation is the calculation of decomposition
coefficients of physical bispinors on basis ones.

The suggested method combines the advantages of both the trace
methods and the computational methods based on spinor technique.
The method as well as the method of spinor techniques enables to
calculate the blocks of the Feynman diagrams (current-like
constructions, spinor products, and by means of the MBS even more
complicated structures) with the help of recursion relation
(\ref{stp19b}) and then to use them in the calculation as
universal functions. The method as well as trace methods does not
require either the compulsory construction of the polarization
vectors of bosons, or the transformation of the slash
construction $\slash{p}$ into bispinors.

The MBS can be easily realized in the systems of symbolic
calculations (Mathematica, Maple, Reduce). As an example, all
calculations have been  done with the help of the simple
rule-based program in environment
``MATHEMATICA''(\cite{Wolfram}). It should be noted that getting
the analytical expression in terms of scalar products on the
given matrix element of the reaction on the ordinary computer
(Pentium-III) takes from $0.2$ to $0.6$ second.

In conclusion, I'd like to emphasize that the purpose of my paper
is to present the new method of analytical calculations of the
matrix element with massless and massive fermions. That is why I
didn't try to stress which method is more efficient (better or
worse), since the criteria of the efficiency can be different.
\appendix
\section{Some relations for massless fermions}
\label{sec:level7}

By means of Eq. (\ref{stp19e}) or (\ref{stp19b}) we can get
expressions for $\Gamma _{C,A;\sigma ,\rho}$ with $Q=\gamma ^\mu
\gamma ^\nu $ and $Q=\gamma ^\mu \gamma ^\nu \gamma ^\alpha $:
\begin{equation}
\Gamma _{C,A;\sigma ,\rho}^{\left\{\mu,\nu\right\}}=2\delta
_{\sigma ,-\rho }\left(
\begin{array}{cc}
\rho \left( b_1^\mu n_{-\rho }^\nu -n_{-\rho }^\mu b_1^\nu \right)
&b_{-1}^\mu b_1^\nu -n_\rho ^\mu n_{-\rho }^\nu  \\
b_1^\mu b_{-1}^\nu -n_{-\rho }^\mu n_\rho ^\nu  & \rho \left(
b_1^\mu n_\rho ^\nu -n_\rho ^\mu b_1^\nu \right)
\end{array}
\right)_{C,A}, \label{app1}
\end{equation}

\begin{eqnarray}
&&\Gamma _{C,A;\sigma ,\rho}^{\left\{\mu,\nu, \alpha
\right\}}=2\delta _{\sigma ,\rho }\left(
\begin{array}{c}
\left( b_1^\mu b_{-1}^\nu -n_\rho ^\mu n_{-\rho }^\nu \right)
b_1^\alpha -\left( n_\rho ^\mu b_1^\nu -b_1^\mu n_\rho ^\nu
\right) n_{-\rho }^\alpha ~~~,
\\
\rho \left( n_\rho ^\mu b_1^\nu -b_1^\mu n_\rho ^\nu \right)
b_{-1}^\alpha +\rho \left( b_1^\mu b_{-1}^\nu -n_\rho ^\mu
n_{-\rho }^\nu \right) n_\rho ^\alpha ,
\end{array}
\right. \nonumber   \\
&&\left.
\begin{array}{c}
\rho \left( n_{-\rho }^\mu b_{-1}^\nu -b_{-1}^\mu n_{-\rho }^\nu
\right) b_1^\alpha +\rho \left( b_{-1}^\mu b_1^\nu -n_{-\rho }^\mu
n_\rho ^\nu
\right) n_{-\rho }^\alpha  \\
\left( n_{-\rho }^\mu b_{-1}^\nu -b_{-1}^\mu n_{-\rho }^\nu
\right) n_\rho ^\alpha +\left( b_{-1}^\mu b_1^\nu -n_{-\rho }^\mu
n_\rho ^\nu \right) b_{-1}^\alpha
\end{array}
\right)_{C,A} .
\label{app2}
\end{eqnarray}

Using  Eq.(\ref{app4}) and the Eqs.(\ref{app1}),(\ref{app2}) we
get two expressions
\begin{eqnarray}
&&M_{\lambda _p,\lambda _k}\left( p,k;\gamma ^\mu \gamma ^\nu
\right)\equiv \overline{U}_{\lambda _p}\left(p\right) \gamma ^\mu
\gamma ^\nu U_{\lambda _k}\left(k\right) = \frac{\lambda _p\delta
_{\lambda _{p},-\lambda _k}}{\sqrt{\left( p\cdot b_{-1}\right)
\left( k\cdot b_{-1}\right) }}
\nonumber \\
&&\left\{ \left( k\cdot n_{\lambda _p}\right) \left[ \left( p\cdot
n_{\lambda _p}\right) \left( b_{-1}^\mu n_{-\lambda _p}^\nu
-n_{-\lambda _p}^\mu b_{-1}^\nu \right) +\left( p\cdot
b_{-1}\right) \left( b_1^\mu b_{-1}^\nu -n_{\lambda _P}^\mu
n_{-\lambda _p}^\nu \right) \right] \right. +
\nonumber \\
&&\left. \left(k \cdot b_{-1}\right) \left[ \left(p \cdot
n_{\lambda _p}\right) \left(n_{-\lambda _p}^\mu n_{\lambda _p}^\nu
-b_{-1}^{\mu} b_1^\nu \right) +\left(p \cdot b_{-1}\right) \left(
n_{\lambda _p}^\mu b_1^\nu -b_1^\mu n_{\lambda _p}^\nu \right)
\right] \right\},
\label{app6}
\end{eqnarray}
\begin{eqnarray}
&&M_{\lambda _p,\lambda _k}\left( p,k;\gamma ^\mu \gamma ^\nu
\gamma ^\alpha \right)  =\frac{\delta _{\lambda _{p},\lambda
_k}}{\sqrt{\left( p\cdot
b_{-1}\right) \left( k\cdot b_{-1}\right) }}  \nonumber \\
&&\left\{ \left( p\cdot b_{-1}\right) \left[ \left( k\cdot
b_{-1}\right) \left( b_1^\alpha \left( b_1^\mu b_{-1}^\nu
-n_{\lambda _p}^\mu n_{-\lambda _p}^\nu \right) +n_{-\lambda
_p}^\alpha \left( n_{\lambda _p}^\mu b_1^\nu
-b_1^\mu n_{\lambda _p}^\nu \right) \right) \right. \right. +
\nonumber \\
&&\left. \left( k\cdot n_{-\lambda _p}\right) \left( n_{\lambda
_p}^\alpha \left( n_{\lambda _p}^\mu n_{-\lambda _p}^\nu -b_1^\mu
b_{-1}^\nu \right) +b_{-1}^\alpha \left( b_1^\mu n_{\lambda
_p}^\nu -n_{\lambda _p}^\mu b_1^\nu
\right) \right) \right] +  \nonumber \\
&&\left( p\cdot n_{\lambda _p}\right) \left[ \left( k\cdot
n_{-\lambda _p}\right) \left( n_{\lambda _p}^\alpha \left(
n_{-\lambda _p}^\mu b_{-1}^\nu -b_{-1}^\mu n_{-\lambda _p}^\nu
\right) +b_{-1}^\alpha \left( b_{-1}^\mu b_1^\nu -n_{-\lambda
_p}^\mu n_{\lambda _p}^\nu \right) \right)
\right. +  \nonumber \\
&&\left. \left. \left( k\cdot b_{-1}\right) \left( b_{-1}^\alpha
\left( b_{-1}^\mu n_{-\lambda _p}^\nu -n_{-\lambda _p}^\mu
b_{-1}^\nu \right) +n_{-\lambda _p}^\alpha \left( n_{-\lambda
_p}^\mu n_{\lambda _p}^\nu -b_{-1}^\mu b_1^\nu \right) \right)
\right] \right\} . \label{app7}
\end{eqnarray}
These relations can be used as universal functions for
calculations of processes with massless fermions.

\newpage
\section*{Figure captions}
\begin{description}
\item{\textbf{Fig.~1}} Feynman diagrams for the process $e^- e^+\to W^- W^+$
\item{\textbf{Fig.~2}} Feynman diagrams for the process $e^- e^+\to f \bar f$
\item{ \textbf{Fig.~3}} One of the possible Feynman diagrams of $e^+e^- \to e^+e^- e^+e^-
e^+e^- $.
\end{description}

\begin{thebibliography}{99}
\bibitem{Powell} J.L. Powell, Phys. Rev. \textbf{75}, ~32~ (1949).
\bibitem{Galynski} M.V. Galynski, S.M. Sikach, hep-ph/9910284.
\bibitem{Bondarev} A.L. Bondarev   hep-ph/9710398.
\bibitem{Farrar} G.R. Farrar  and F. Neri, Phys. Lett. \textbf{130B}, ~109~(1983).
\bibitem{Giele} F.A. Berends  and  W.T. Giele, Nucl. Phys. \textbf{B294}, 700~(1987).
\bibitem{Caravaglios}F. Caravaglios,  M. Moretti, Phys. Lett. \textbf{B358}, 332~ (1995).
\bibitem{Helac}A. Kanaki and  C.G. Papadopoulos,
Comput. Phys. Commun. \textbf{132}, 306~ (2000).
\bibitem{Bellomo}E. Bellomo, I1 Nuovo Cimento. Ser.X \textbf{21}, 730~ (1961).
\bibitem{Bogush} A.A. Bogush, F.I. Fedorov, ~Vesti AN BSSR, ser.fiz.-m.n.
\textbf{N~2}, 26~(1962) (in Russian).
\bibitem{Bogush1} A.A. Bogush, ~Vesti AN BSSR, ser.fiz.-m.n., \textbf{N~2}, 29~ (1964)
(in Russian).
\bibitem{Bjorken}J.D. Bjorken  and M.C. Chen, Phys.Rev.  \textbf{154}, 1335~ (1966).
\bibitem{Fearing} H.W. Fearing, R.R. Sillbar, Phys.Rev. \textbf{D6}, 471~ (1972).
\bibitem{Fedorov} F.I. Fedorov, Izvestiya Vyzov. Fizika,  \textbf{N~2}, 32~
(1980) (in Russian).
\bibitem{Caffo}M. Caffo, E. Remiddi,  Helvetica Phys. Acta. \textbf{55}, 339~ (1982)
\bibitem{Vega}R. Vega  and J. Wudka, Phys. Rev. \textbf{D53}, 5286 (1996);
\textit{erratum}  Phys. Rev. \textbf{D56}, 6037~(1997).
\bibitem{Berendz} F.A. Berendz, P.H. Davervedt, and  R. Kleiss,
Nucl. Phys. \textbf{B253}, 141~ (1985).
\bibitem{Kleiss} R. Kleiss, W.J. Stirling, Nucl. Phys. \textbf{B262}, 235~(1985).
\bibitem{Zhan}Zhan Xu, Da-Hua Zhang, and Lee Chang, Nucl.Phys. \textbf{B291},
392~(1987).
\bibitem{Kleiss1}R. Kleiss, Z.Phys.C. \textbf{33}, 433~(1987).
\bibitem{Hagiwara}K. Hagiwara,  D. Zeppenfeld,  Nucl. Phys. \textbf{B274}, 1~(1986).
\bibitem{Gongora}T. A.-Gongora, R.G. Stuart,  Z.Phys. \textbf{C42},
617~(1989).
\bibitem{Andreev}V.V. Andreev,  Phys.Rev. \textbf{D62},~142029~(2000).
\bibitem{Papadopoulos} E.N. Argyres, C.G. Papadopoulos, Phys. Lett. \textbf{B263}, 298~(1991).
\bibitem{Ballestrero} A. Ballestrero and E. Maina, Phys. Lett. \textbf{B350}, 225~ (1995).
\bibitem{Comphep}A.E. Pukhov et al., Preprint INP MSU 98-41/542 (1998); hep-ph/9908288.
\bibitem{Grace} KEK Report 92-19, February (1993).
\bibitem{FeynCalc}R. Mertig, M.B\"ohm, and  A. Denner, Comp. Phys.
Commun. \textbf{64}, 345~(1991). .
\bibitem{Stelzer}T. Stelzer  and  W.F. Long, Comput. Phys. Commun. \textbf{81}, ~357~(1994).
\bibitem{Harlander}R. Harlander  and M. Steinhauser,   Preprint TTP98-41, BUTP-98/28 1998. ;
hep-ph/9812357.
\bibitem{Bjorken1}J.D. Bjorken and S.D. Drell, Relativistic
Quantum Mechanics. McGraw-Hill, New York, (1964).
\bibitem{Borodulin}V.I. Borodulin, R.N. Rogalyov, and S.R. Slabospitsky, Preprint IHEP 95-90, (1995); hep-ph/9507456.
\bibitem{Ilyin} P.S. Cherzor, V.A. Ilyin, A.E.  Pukhov,  hep-ph/0101265.
\bibitem{Jadach} S. Jadach, B.F.L. Ward, and Z. Was, Phys. Lett. \textbf{B449}, 97~(1999).
\bibitem{Hagiwara1}K. Hagiwara, R.D. Peccei, and D. Zeppenfeld,
Nucl.Phys. \textbf{B282}, 253~ (1987).
\bibitem{Wolfram} St. Wolfram, The Mathematica book. 4th edition. Addison-Wesley, (1999).
\end{thebibliography}
\end{document}